\documentclass[%
aps,
reprint,
superscriptaddress,
amsmath,amssymb,
prb,
floatfix,
]{revtex4-1}

\usepackage{chemformula}
\usepackage[utf8]{inputenc}
\usepackage[T1]{fontenc}
\usepackage{graphicx}
\usepackage{epstopdf}
\usepackage{siunitx}
\usepackage{amsmath}
\usepackage{amssymb}
\usepackage[colorlinks=true,allcolors=blue]{hyperref}
\usepackage{soul}
\usepackage[normalem]{ulem}

\begin{document}


\title{Dual Topological Insulator Device with Disorder Robustness} 

\author{Bruno Focassio}\email{b.focassio@ufabc.edu.br}
\affiliation{Center for Natural and Human Sciences, Federal University of ABC (UFABC), 09210-580, Santo André , São Paulo, Brazil}
\affiliation{Brazilian Nanotechnology National Laboratory (LNNano), CNPEM, 13083-970, Campinas, São Paulo, Brazil}

\author{Gabriel R. Schleder}
\affiliation{Center for Natural and Human Sciences, Federal University of ABC (UFABC), 09210-580, Santo André , São Paulo, Brazil}
\affiliation{Brazilian Nanotechnology National Laboratory (LNNano), CNPEM, 13083-970, Campinas, São Paulo, Brazil}

\author{Armando Pezo}
\affiliation{Center for Natural and Human Sciences, Federal University of ABC (UFABC), 09210-580, Santo André , São Paulo, Brazil}
\affiliation{Brazilian Nanotechnology National Laboratory (LNNano), CNPEM, 13083-970, Campinas, São Paulo, Brazil}

\author{Marcio Costa}
\affiliation{Physics Department, Fluminense Federal University, 24210-346, Niterói, Rio de Janeiro, Brazil}
\affiliation{Brazilian Nanotechnology National Laboratory (LNNano), CNPEM, 13083-970, Campinas, São Paulo, Brazil}

\author{Adalberto Fazzio}\email{adalberto.fazzio@lnnano.cnpem.br}
\affiliation{Brazilian Nanotechnology National Laboratory (LNNano), CNPEM, 13083-970, Campinas, São Paulo, Brazil}
\affiliation{Center for Natural and Human Sciences, Federal University of ABC (UFABC), 09210-580, Santo André , São Paulo, Brazil}

\date{\today}

\begin{abstract}
Two-dimensional \ch{Na3Bi} is a dual topological insulator protected by time-reversal and mirror symmetry, resulting in a promising platform for devices design.
However, in reality, the design of topological devices is hindered by a sensitivity against disorder and temperature. 
We study the topological properties of \ch{Na3Bi} in the presence of intrinsic defects, investigating the robustness of the edge states and the resulting transport properties. We apply a recursive Green's function technique enabling the study of disordered systems with lengths comparable to experimentally synthesized materials, in the order of micrometers. We combine our findings to propose a topological insulator device, where intrinsic defects are used to filter the response of trivial bulk states. This results in a stable conductance throughout a large range of electronic temperatures, and controllable by a perpendicular electric field.
Our proposal is general, enabling the design of various dual topological insulators devices.
\end{abstract}

\maketitle 


\section{Introduction}

\citet{qshi_graphene_KaneMele} showed that the spin-orbit interaction plays a key role in the topology of the electronic structure of a graphene lattice, realizing the quantum spin Hall (QSH) state that is characterized by a $\mathbb{Z}_2$ topological invariant \cite{z2_KaneMele.95.146802}. Distinguished by the non-zero $\mathbb{Z}_2$ invariant, quantum spin Hall insulators (QSHIs) feature counter-propagating helical gapless states localized at the boundary with a trivial insulator \cite{HansanKane_colloquium,review_tis,ando_review,band_inversion}. These edge (surface) states are robust against elastic backscattering due to the bulk electronic structure protected by time-reversal symmetry (TRS) \cite{HansanKane_colloquium,review_tis,ando_review,band_inversion} and the bulk boundary equivalence \cite{bulk_boundary}. The presence of metallic edge states is mainly characterized by angle-resolved photoemission spectroscopy (ARPES), probing the electronic structure at the edges (surfaces) of a 2D (3D) QSHI \cite{hgte_quantum_wells,Bernevig_quantum_well}. To date, there are few QSHIs experimentally realized and several theoretically proposed \cite{hgte_quantum_wells,quantum_well_2,tmdcs_qshi_device,bi_monolayers,amorphous_TI,review_dft_to_ml,TIs_ML,marzari_new_TI}, but their characterization and application is often limited by the size of the bandgap, which is usually small \cite{ando_review,band_inversion}.

Besides TRS, crystalline symmetries may also protect a distinct topological character, resulting in topological crystalline insulators (TCIs) \cite{Fu_TCI,ando_tci}. The TCI state is more commonly found to be protected by mirror symmetries, and also results in dissipationless states at the boundary with a trivial insulator. Mirror protected TCIs are characterized by a non-zero mirror Chern number ($\mathcal{C}_M$) \cite{mirror_chern_number}. Several other symmetries may protect the bulk topology resulting in different topological insulators (TIs) \cite{HOTI_1,HOTI_2,floquet_tis,particle_hole_ti}, a particular and yet less explored case is the dual topological insulator (DTI),  when the bulk topology is simultaneously protected by TR and crystal symmetries \cite{mirror_chern_number,experiment_dtc_1,experiment_dtc_2}. For QSHIs (TCIs), breaking TR (crystalline) symmetry would result in a trivial insulator. As a DTI, breaking time-reversal (mirror) symmetry alone would still result in gapless edge states protected by the mirror (time-reversal) symmetry, allowing us to separately control both topological phases \cite{na3bi_carlos}.

\ch{Na3Bi} is a DTI among those with the largest bandgap, additionally, it also possesses a robust band inversion and great stability under strain \cite{blugel_na3bi}. Recently, an electrical driven quantum phase transition was experimentally observed~\cite{Efield_na3bi_nature}, turning \ch{Na3Bi} in a promising candidate to be applicable on TI devices. The experimental realization of \ch{Na3Bi} and its characterization revealed the presence of spontaneous occurring defects at step edges that could alter the material properties in the application~\cite{Efield_na3bi_nature}. However, the investigation of the electronic structure and transport properties of defective DTIs is unexplored in the literature. 


We study \ch{Na3Bi} defects at the edges of armchair nanoribbons. We employ density functional theory (DFT) \cite{dft1964,dft1965} for geometry optimization and electronic structure calculations, obtaining the \textit{ab initio} Hamiltonians for transport calculations. We make use of the non-equilibrium Green's function (NEGF)\cite{transport_datta_book,transport_1_Caroli1971,transport_datta_paper,transport_basic_theory_2,transport_basic_theory_3_sanvito,transport_basic_theory_4_nardelli} method to account for transport properties and using a recursive scheme we study the effect of disorder on \ch{Na3Bi} with ribbon lengths up to \SI{0.2}{\micro\metre}. Inspired by the electric field topological phase transition, we propose the use of disordered \ch{Na3Bi} to compose a device that is highly controllable and stable at high electronic temperatures.

\section{Methods}

\subsection{Electronic structure}

Bulk electronic structure calculations shown in Fig. \ref{fig:results:structure_and_bands} were performed with  VASP~\cite{vasp1,vasp2} using the projector augmented-wave (PAW) method~\cite{paw} and self-consistent spin-orbit coupling (SOC). The kinetic energy cutoff for the plane wave expansion was \SI{520}{\electronvolt}, with k-point density of 20 $/$\si{\per\angstrom}.

Nanoribbons electronic structures and defects relaxations, shown in Fig. \ref{fig:results:defects}, were performed with the SIESTA~\cite{siesta_code} code using an optimized polarized single-$\zeta$ (SZP) basis-set, which is sufficient to replicate the plane-waves band structure with satisfactory accuracy.
The geometry optimization procedure for the considered defects was performed until the Hellmann-Feynman forces were smaller than \SI{e-2}{\electronvolt\per\angstrom}. The electronic structure calculations were performed with the on-site self-consistent SOC~\cite{siesta_soc_1,siesta_soc_2} as well as fully-relativistic norm-conserving pseudopotentials~\cite{tm_pseudopotentials} and energy grid cut-off of 350 Ry. As for k-space sampling, we used a k-point density larger than 16 $/$\si{\per\angstrom}. We added \SI{15}{\angstrom} of vacuum to avoid spurious interactions between periodic images in non-periodic directions. We passivated the ribbons' edges with hydrogen to circumvent dangling bonds effects. Furthermore, the pristine leads are modeled as one unit-cell armchair nanoribbons with \SI{9.22}{\angstrom} in length, and the scattering region as four unit-cell armchair nanoribbons with \SI{36.89}{\angstrom} in length. All nanoribbons are \SI{85.90}{\angstrom} wide, which is sufficient to present gapless edge states.

\subsection{Ballistic transport}

\begin{figure}
	\centering
	\includegraphics[width=\linewidth]{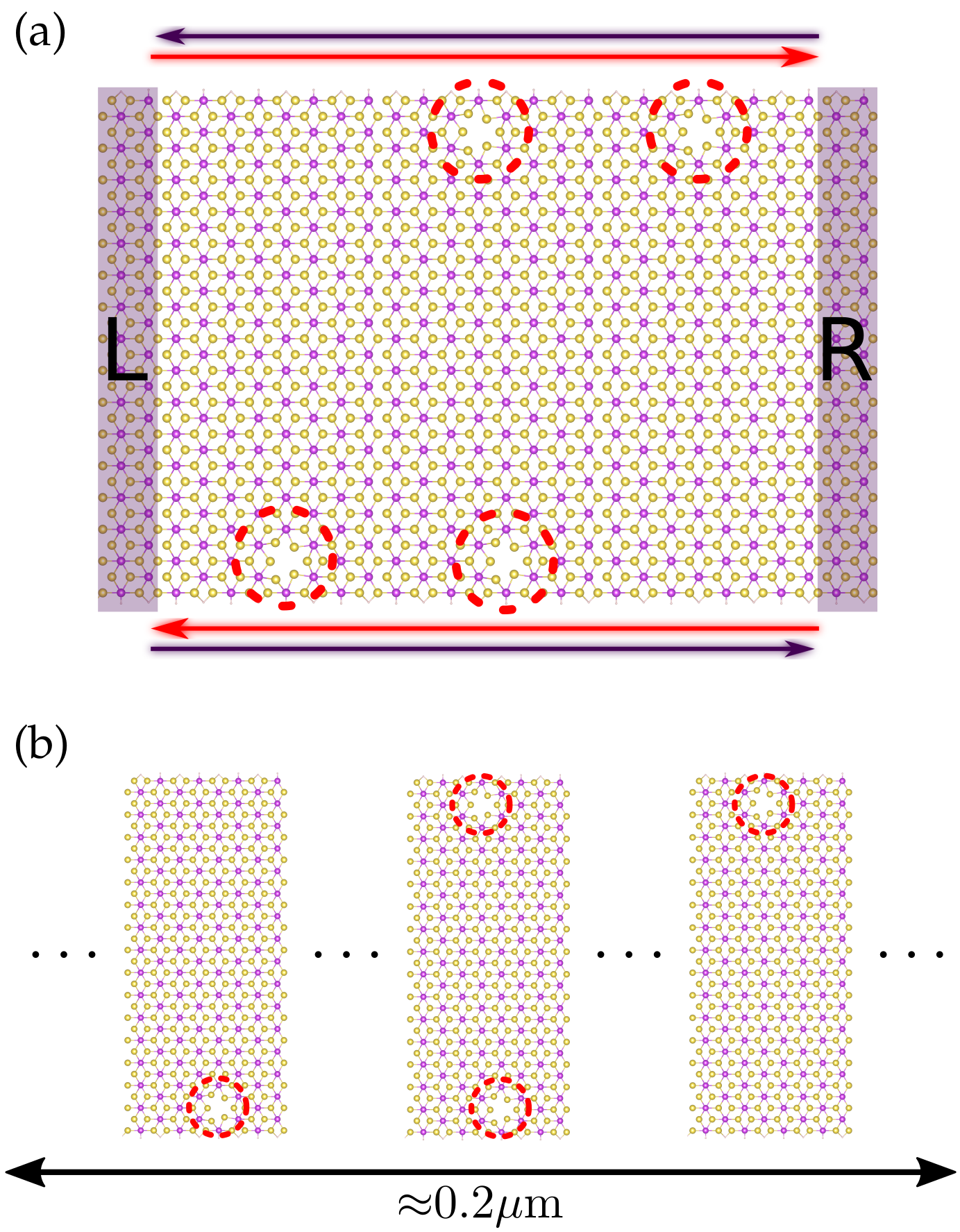}
	\caption{Representation of the Na$_3$Bi electronic device setup composed of several building blocks. (a) Device with left (L) and right (R) leads, showing the transport direction of spin currents. (b) Scheme of central region composed of different building blocks, represented pictorically with distance between each block. A device with 50 building blocks reaches approximately \SI{0.2}{\micro\metre}.}
	\label{fig:device_example}
\end{figure}

The transport coefficients were obtained using the system's setup with left (L) and right (R) semi-infinite leads with a scattering region (S) in between, see Fig. \ref{fig:device_example}(a). In the small bias regime and at \SI{0}{\kelvin}, the conductance through the scattering region can be expressed by the Landauer formula $G = (e^2/h) T(\varepsilon_F) $~\cite{landauer_formula}, where $T(\varepsilon_F)$ is the transmission at the Fermi level. The transmission is written in terms of the Green's function of the scattering region $G_S$ and the coupling matrices between the scattering region and the leads $\Gamma_{L,R}$ as in Eq. \eqref{eqn:methods:transmission} \cite{transport_1_Caroli1971,trace_formula}. To compute the scattering region Green's function and the leads coupling matrices we implemented the method described by ref. \cite{transport_basic_theory_3_sanvito,Rocha2006,transport_decimation_theory} using the self-consistent converged DFT Hamiltonians obtained directly from SIESTA.
 
\begin{equation}
    T(E) = \text{Tr}[\Gamma_L G_S \Gamma_R G_S^{\dagger}] \label{eqn:methods:transmission}
\end{equation}

The method itself does not limit the size of the scattering region, the main difficulty arises in the calculation of the Green's function using an \textit{ab initio} Hamiltonian, which may be challenging regarding the system's size and the basis set used in the calculation. To achieve real length scattering regions and to enable the study of disordered systems we must combine the NEGF method with the recursive Green's function technique~\cite{transport_basic_theory_3_sanvito,transport_decimation_theory,decimation_caio,decimation_3}.
The recursive procedure starts by splitting the total scattering region into building blocks in such a way that each building block has an isolated defect, as illustrated in Fig. \ref{fig:device_example}(b). The blocks are connected via first neighbors' interactions. This division of the scattering region allows us to compute the \textit{ab initio} DFT Hamiltonian and overlap matrix for each building block separately and use the method to recursively eliminate internal degrees of freedom of the scattering region Hamiltonian. The result is an effective Hamiltonian describing the whole scattering region. The computational cost of using this technique to compute the system's transport coefficients is similar to the computational cost of calculating the Green's function of a single building block. 
As such, the recursive method has already been successfully applied to study several large length systems, reaching scales up to hundreds of nanometers \cite{transport_decimation_theory,decimation_caio,decimation_3,decimation_application_1,decimation_application_2,decimation_application_3,decimation_pezo_butadiene}.

\section{Results and Discussion}

\begin{figure}[!htb]
    \centering
    \includegraphics[width=\linewidth]{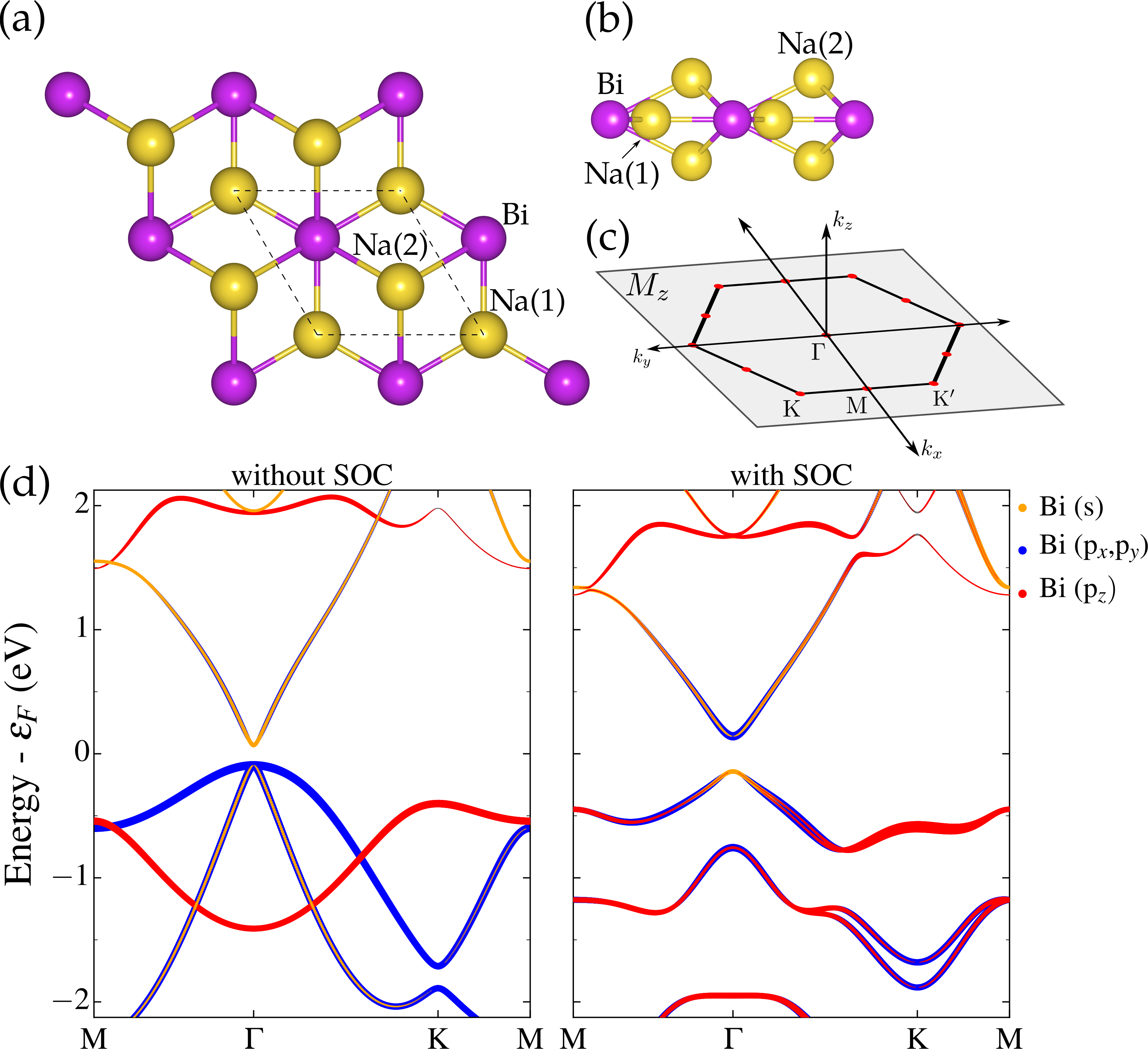}
    \caption{\ch{Na3Bi} structure and electronic band structure. (a) Atomic structure showing Bi, Na(1) in-plane atoms, and Na(2) out-of-plane atoms, along with the unit cell in dashed lines. (b) Side view of the atomic structure. (c) Brillouin Zone indicating high-symmetry points and mirror plane $M_z$. (d) Orbital projected band structure without and with spin-orbit coupling (SOC).
    }
    \label{fig:results:structure_and_bands}
\end{figure}

Our starting reference is bulk \ch{Na3Bi}, with its two-dimensional atomic arrangement depicted in Fig.~\ref{fig:results:structure_and_bands}. Sodium atoms have two distinct positions, Na(1) as in-plane atoms and Na(2) as out-of-plane atoms. \ch{Na3Bi} forms a honeycomb lattice preserving three-fold rotation about the z-axis as shown in Fig.~\ref{fig:results:structure_and_bands}(a). Besides $R_3$ symmetry, there is a mirror-symmetric plane $M_z$ perpendicular to the z-axis ($z \rightarrow -z$), as illustrated in Fig.~\ref{fig:results:structure_and_bands}(b)--(c), that plays a key role in the topology of \ch{Na3Bi} electronic structure.
The 3D counterpart of \ch{Na3Bi} is composed of stacked monolayers and is a topological Dirac semimetal (TDS) \cite{discovery_na3bi_tds}, in contrast, 2D \ch{Na3Bi} is an insulator~\cite{blugel_na3bi}. Figure \ref{fig:results:structure_and_bands}(d) shows its band structure with and without spin-orbit coupling (SOC). Without SOC, the bandgap is \SI{0.16}{\electronvolt} located at $\Gamma$, with the valence band (VB) and conduction band (CB) mainly composed of Bi-$p_x,p_y$ orbitals with a small contribution of Bi-$s$ orbital at the CB minimum (CBM). With SOC, there is a band inversion at $\Gamma$, with the VB mainly composed of Bi-$p_x,p_y$ with a larger contribution of  Bi-$s$ at the VB maximum (VBM), and the CBM switches to mainly Bi-$p_x,p_y$ orbitals. The bandgap with SOC is \SI{0.30}{\electronvolt} and it remains direct at $\Gamma$, in agreement with previous results \cite{blugel_na3bi,Efield_na3bi_nature}.
\ch{Na3Bi} is a well known DTI, with its dual topological character due to time-reversal and crystalline $M_z$ symmetry \cite{blugel_na3bi,Efield_na3bi_nature}. We verified \ch{Na3Bi} topology computing the $\mathbb{Z}_2$ and $\mathcal{C}_M$ topological invariants \cite{z2pack_1,z2pack_2,mirror_chern_number}, which resulted in $1$ and $-1$, respectively.

\begin{figure}[!htb]
    \centering
    \includegraphics[width=\linewidth]{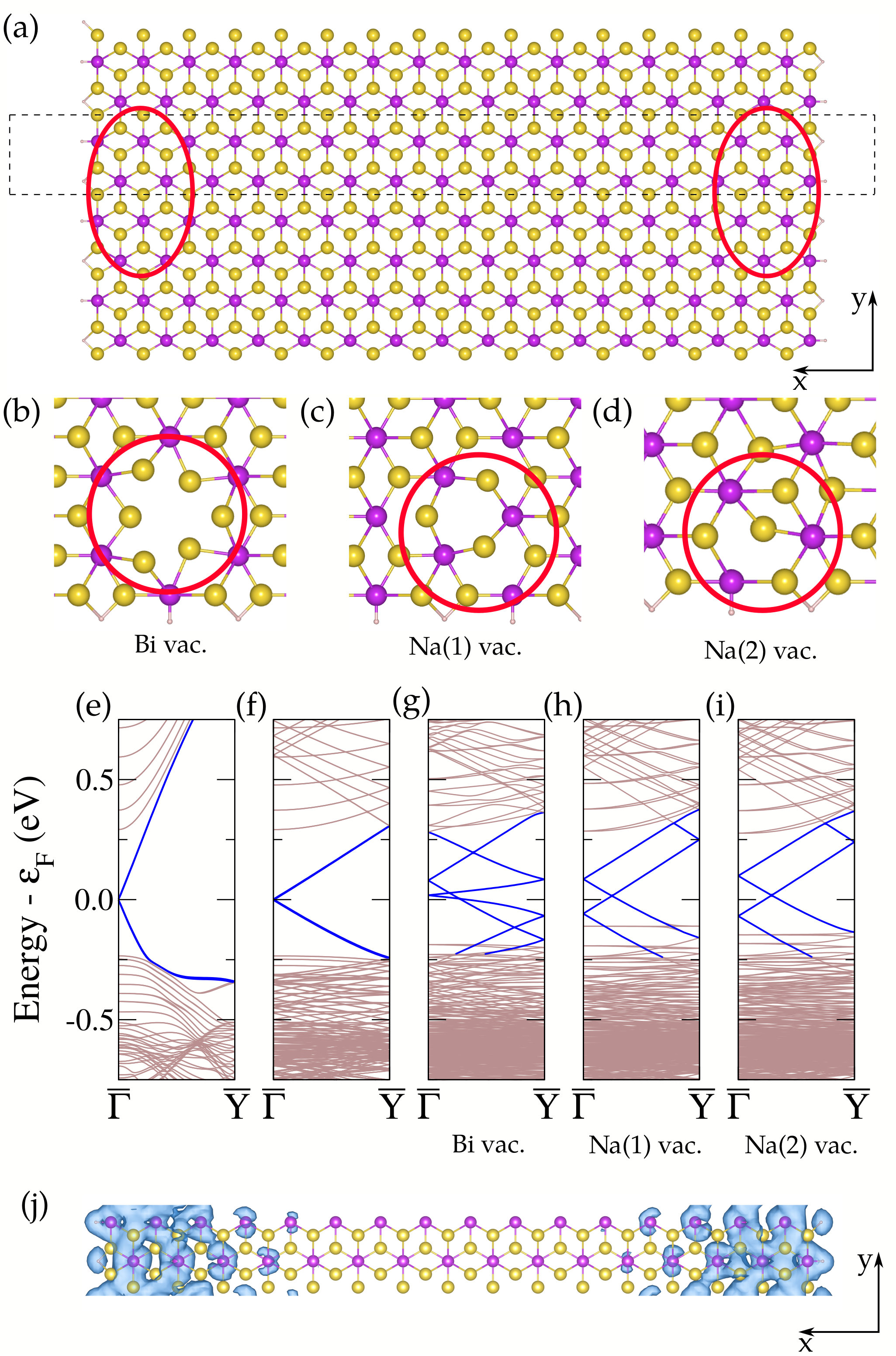}
    \caption{(a) \ch{Na3Bi} 4 unit-cells nanoribbon, with unit-cell marked in dashed lines. (b) Bi vacancy. (c) Na(1) vacancy. (d) Na(2) vacancy. Band structure of (e) 1 unit-cell nanoribbon, (f) 4 unit-cells nanoribbon, and nanoribbon with (g) Bi vacancy, (h) Na(1) vacancy, (i) Na(2) vacancy. We explored these defects on each and both edges. (j) Charge density of  states at Fermi level in panel (e); isosurface level of \SI{0.03}{\per\cubic\angstrom}.}
    \label{fig:results:defects}
\end{figure}

The interface between topological and trivial insulators carries gapless states that are robust to disorder as long as the bulk boundary correspondence holds and the edge preserves the same symmetry that protects the bulk topology~\cite{bulk_boundary}. To access such robustness we study intrinsic defects near \ch{Na3Bi} edges. 
First, we compute the band structure (Fig.~\ref{fig:results:defects}(e)) of a \ch{Na3Bi} armchair nanoribbon shown in Fig.~\ref{fig:results:defects}(a). The ribbon band structure clearly shows gapless states inside the bulk bandgap, and in Fig. \ref{fig:results:defects}(j) the charge density plot verifies that these states are spatially located at the ribbon's edges. We repeated the unit-cell armchair ribbon along the periodic direction to create a four-unit-cell ribbon to be used as a platform for defect calculations. The geometry and band structure are seen in Fig.~\ref{fig:results:defects}(a) and \ref{fig:results:defects}(f), respectively. Next, we place vacancies close to the edges of these ribbons (positions marked in red in Fig.~\ref{fig:results:defects}(a)).

Na(2) vacancies with formation energy of \SI{0.7}{\electronvolt} should be much more common than both Na(1) and Bi vacancies, with formation energies of \SI{0.9}{\electronvolt} and \SI{1.9}{\electronvolt}, respectively. As observed in experiments \cite{Efield_na3bi_nature}, Na(2) vacancies are expected to appear at the edges. With a Bi vacancy, Fig. \ref{fig:results:defects}(g), the dispersion of the edge state is decreased, but remains linear near $\Gamma$ with the bandgap closed and the folded topological states emerging from the bulk bands. The edge states of Na(1) and Na(2) defects (Fig.~\ref{fig:results:defects}(h) and \ref{fig:results:defects}(i), respectively) affect only the defective edge also breaking edge degeneracy. We observe non-topological defect-like localized states inside the gap near the valence band which intersects the edge states.

The presence of these defects locally breaks the mirror symmetry as a result of relaxation, for Na(1) and Bi defects, and also by the missing atomic position for Na(2) defects. There is no net magnetic moment resulting from these defects, preserving TRS. The presence of defects on TIs can cause interference on edge states as a result of small ribbon widths~\cite{thygesen_transport} or high vacancy concentration~\cite{vacancies_TI_transport_Tiwari_2019}, where the defects mediate the coupling between the states on opposite edges leading to backscattering. The ribbons used in our calculations are almost \SI{86}{\angstrom} wide, therefore, the defects studied at concentrations of about \SI{0.4}{\percent} are not enough to disturb the correspondence with the bulk and they do not interfere with the topological states, remaining with a closed bandgap. In the following, we show that edge defects besides preserving topological states, are also beneficial for transport properties.

\begin{figure}[!htb]
    \centering
    \includegraphics[width=\linewidth]{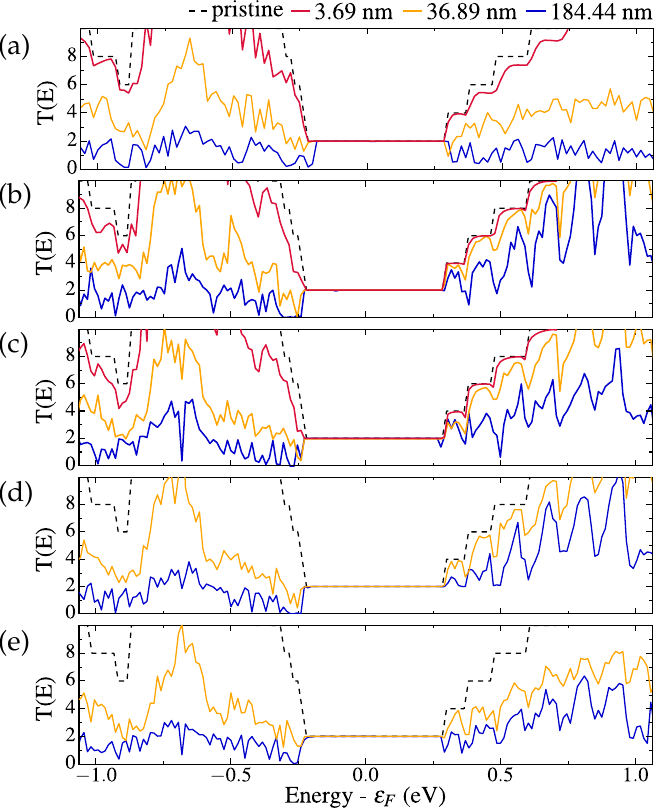}
    \caption{Electronic transport transmission spectrum for defected \ch{Na3Bi} at various lengths, mixing defects in one and both edges. (a) Bi vacancies. (b) Na(1) vacancies. (c) Na(2) vacancies. (d) Na(1) and Na(2) vacancies. (e) Bi and both Na vacancies.}
    \label{fig:results:transmission_defects}
\end{figure}

To access the robustness of \ch{Na3Bi} against disorder we compute the total electronic transmission of the nanoribbons through the recursive technique considering different degrees of disorder. In Fig.~\ref{fig:results:transmission_defects} we consider mixed defects on either and both edges, starting from a single defect (\SI{3.69}{\nano\metre}) until reaching a scattering region with \SI{184.44}{\nano\metre}. 
Figure ~\ref{fig:results:transmission_defects}(a) demonstrates the effect of Bi defects solely. Bulk states are localized by the defects resulting in the decrease of transmission as the ribbon length increases, however, the edge states are little affected by the presence of defects and the transmission remains quantized at the bandgap even with long ribbon lengths. Figure~\ref{fig:results:transmission_defects}(b)--(d) shows the transmission for Na defects. Na(1) and Na(2) (see Fig.~\ref{fig:results:transmission_defects}(b) and ~\ref{fig:results:transmission_defects}(c), respectively) defects do not change the transmission of the states inside the gap, although they hinder the bulk transmission. Mixing Na(1) and Na(2) defects (see Fig.~\ref{fig:results:transmission_defects}(d)) shows the same trend, increasing the ribbon length only modifies the bulk transmission. Also, as seen in Fig.~\ref{fig:results:transmission_defects}(e), adding Bi defects to the scattering region has a greater localization effect for bulk trivial states than Na defects alone.

Figure \ref{fig:results:transmission_defects} provides an insight into the robustness of the edge states of a topological insulator under structural disorder. Even with defects that locally break the mirror symmetry, the topological states proved to be highly robust. The recent realization of an amorphous TI~\cite{amorphous_TI} proved that topological states may be persistent in strong structural disorder regimes. 
In SiC supported bismuthene, the topological states resist a random vacancy concentration threshold of almost \SI{17}{\percent}~\cite{robustness_ti_vacancies}, with this value dependent on the material's band gap and SOC strength \cite{robustness_ti_vacancies,vacancies_TI_transport_Tiwari_2019}. In \ch{Na3Bi}, with vacancy concentration up to \SI{0.4}{\percent}, the transmission (conductance) of the topological edge states take a constant value of 2 (G$_0$) resulting from the edge states dispersion even in length scales of micrometers. Figure \ref{fig:results:transmission_defects} clearly shows that bulk states transmission are vanishingly small in real length scattering regions, therefore, enabling the construction of devices where defects are desirable, allowing the filtering of bulk states while retaining only the response of topological states.

\begin{figure}[!htb]
	\centering
	\includegraphics[width=\linewidth]{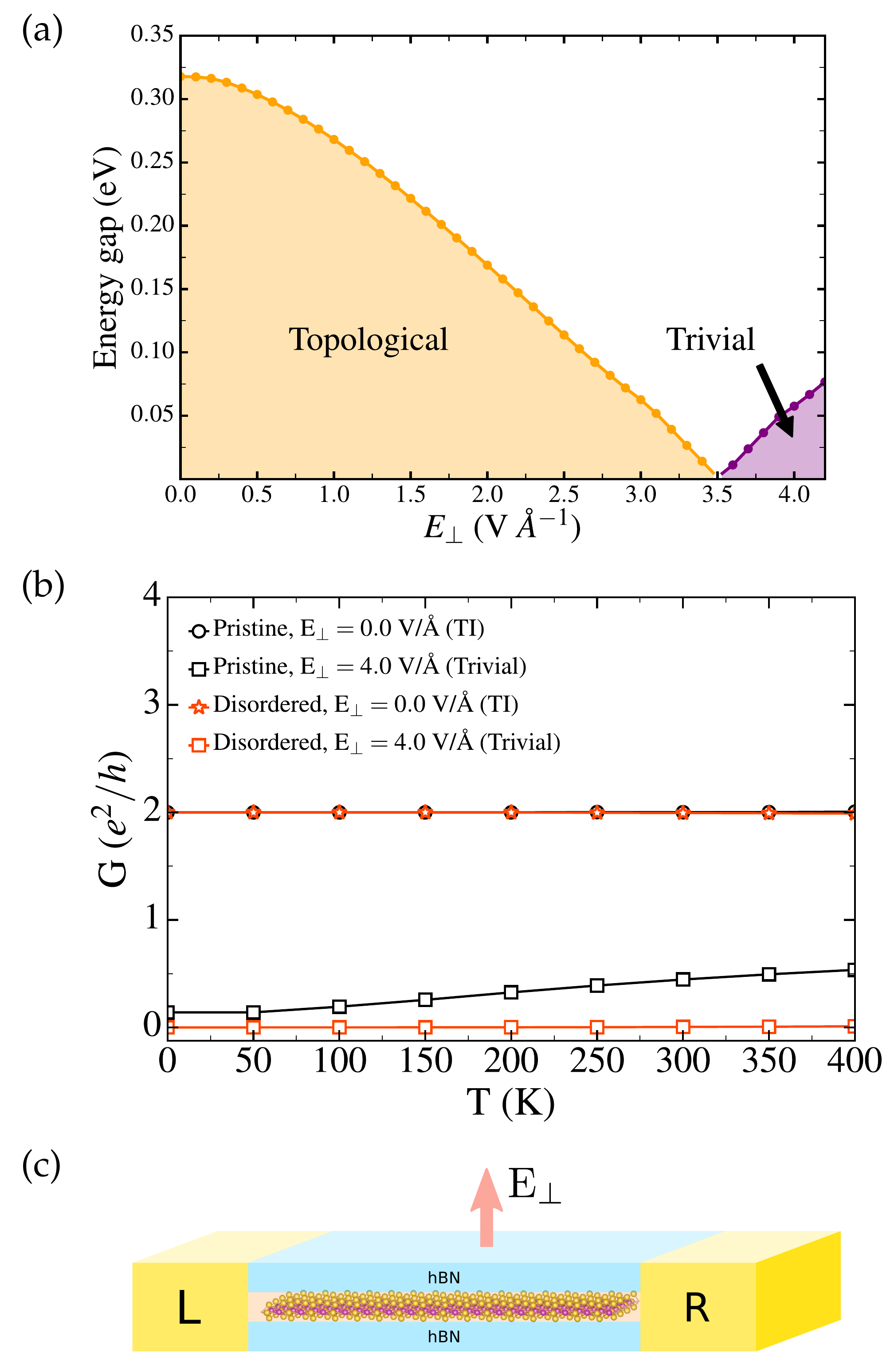}
	\caption{Electric field control of topological phase transition, effect on differential conductance with electronic temperature and proposed device. (a) Electric field driven topological phase diagram obtained in SIESTA, with the topological phase transition value of approximately $E_\perp = \SI{3.5}{\volt\per\angstrom}$. (b) Differential conductance as a function of electronic temperature for both pristine (black) and disordered (with Bi defects; orange) \ch{Na3Bi}, with and without the application of a perpendicular \SI{4.0}{\volt\per\angstrom} electric field. The device length is \SI{184.44}{\nano\metre}. (c) Representation of a DTI device composed of disordered \ch{Na3Bi}, controllable by the perpendicular electric field, and encapsulated by hBN.}
	\label{fig:results:field}
\end{figure}

Topological insulators devices should possess two main features allowing their application: controllability, with the switch between on and off states, and thermal stability. \ch{Na3Bi} has a topological phase transition induced by a perpendicular electric field, as displayed in Fig. \ref{fig:results:field}(a). The on and off switch is performed by the conductance having a value of $2e^2/h$ and zero, respectively, see Fig. \ref{fig:results:field}(b). Electronic and structural temperature effects can give rise to the contribution of bulk states to the conductance, resulting in a higher conductance and reducing the device's controllability \cite{temperature_response_TI}, as displayed in Fig. \ref{fig:results:field}(b).

Using the effect caused by defects, we propose a TI device  controllable through a topological phase transition, and with electronic thermal stability at high electronic temperatures, see Fig. \ref{fig:results:field}(c). The main feature here is the use of defects to filter the bulk states stabilizing the response of the device. 
We investigate the phase transition produced by an electric field, that is a result of breaking the mirror symmetry, and also the Stark effect that rearranges the bands orbital character \cite{Efield_na3bi_nature}. The control of topological states through the application of an electric field has been proposed in several works \cite{blugel_na3bi,tmdcs_qshi_device,phosphorene_electric_field}, but the consequences in transport properties in disordered TIs, and DTIs, and the effect of electronic temperature in these systems still lacks a proper investigation.

In the linear regime, one can compute the differential conductance $G$ using Eq. \eqref{eqn:methods:conductance} \cite{transport_datta_book}.

\begin{equation}
G = \frac{e^2}{h} \int T(E) \left( -\frac{\partial f(E)}{\partial E} \right) \mathrm{d}E \label{eqn:methods:conductance}
\end{equation}

\noindent where the derivative of the Fermi-Dirac distribution is the thermal broadening function accounting for electronic temperature effects in the conductance.

In Fig.~\ref{fig:results:field}(b), we evaluate Eq. \eqref{eqn:methods:conductance} as a function of the temperature, as given by the derivative of the Fermi-Dirac distribution, for four different configurations: (i) pristine \ch{Na3Bi}, (ii) pristine \ch{Na3Bi} with a perpendicular electric field applied, (iii) disordered \ch{Na3Bi} with no applied electric field and (iv) disordered \ch{Na3Bi} with a perpendicular electric field applied ($E_{\perp} =\;$\SI{4.0}{\volt\per\angstrom}).

The large band gap in \ch{Na3Bi} ensures a quantized conductance for medium temperatures, regardless of the presence of structural disorder. However, with the applied electric field, as the electronic temperature increases, pristine \ch{Na3Bi} shows an increase in conductance due to tunneling between states of the valence and conduction bands, harming the on/off ratio, and therefore the controllability. The same is not observed for disordered \ch{Na3Bi}, the conductance vanishes with the application of a perpendicular electric field in the disordered regime; the conductance is close to zero even at high temperatures, resulting in a thermally stable response. Even though we only include electronic temperature effects, \ch{Na3Bi} has shown to be highly stable against strain up to 20\% change in the lattice parameter without closing the band gap~\cite{blugel_na3bi}, thus we anticipate that thermal expansion should not play a major role in \ch{Na3Bi} transport properties.

For TIs with a smaller band gap than \ch{Na3Bi}, the cleansing of topological response due to disorder should be much more pronounced, deeply affecting the transport properties. For instance, \ch{WTe2} is experimentally reported to show a similar behavior of the conductance with temperature, but with an increase in conductance around \SI{100}{K}~\cite{temperature_response_TI}, demonstrating that our conclusions are not limited to \ch{Na3Bi} nor DTIs, but are valid for topological insulators in general.

\section{Conclusions}

In summary, we investigate \ch{Na3Bi} topological properties focusing on vacancy edge defects in real length nanoribbons. We study the electronic structure of the edge with several vacancy defects configurations on large nanoribbons. We find that locally breaking the bulk symmetry does not eliminate the correspondence between bulk and boundary, ensuring topological edge states for defective nanoribbons. 
Using a recursive Green's function technique, we investigate the role of defects on transport properties of \ch{Na3Bi} ribbons with lengths up to \SI{0.2}{\micro\metre}. Our transport calculations show that even at a high number of vacancy defects, we still observe a quantized conductance at the bandgap. 
Additionally, there is a filtering effect on bulk states whose conductance contribution vanishes at high disorder regimes, granting only the topological response. This effect can be even more important for other TIs with smaller band gaps. 
Finally, we show that with the application of an electric field, we can switch between a topological and trivial insulator, and employing this topological phase transition we propose a DTI device enhanced by disorder filtering that is highly controllable and stable.
While we show this proposal using the specific \ch{Na3Bi} system, our proposal is general in the sense that it enables the design of dual topological insulators devices of several different DTI materials.

\begin{acknowledgments}

This work is supported by São Paulo Research Foundation (FAPESP), grants no. 19/04527-0, 16/14011-2, 17/18139-6, and 17/02317-2, and the Brazilian National Council for the Improvement of Higher Education (CAPES). The authors acknowledge the Brazilian Nanotechnology National Laboratory (LNNano/CNPEM, Brazil) and the Santos Dumont supercomputer at the Brazilian National Scientific Computing Laboratory (LNCC) for computational resources.

\end{acknowledgments}

\bibliographystyle{apsrev4-2}
\bibliography{main}

\end{document}